# Comparative tests of accretion rates of quasars derived using a new empirical and an existing theoretical relation: Insights into black hole properties and growth

Yash Aggarwal

Emeritus Associate: Lamont-Doherty Earth Observatory
Palisades, NY 10965, USA
Current address: 822 Winton Drive, Petaluma, CA 94954
E-mail: haggarwal@hotmail.com

## ABSTRACT

A scaling relation based on thin-disc accretion theory has been used by some workers to determine the mass-inflow rate $\dot{M}_{disk}$ onto 20 high-redshift (z>5.3) and 80 Palomar-Green quasars at z<0.5. Based on several assumptions, it inexplicably implies that $\dot{M}_{disk}$ is an inverse function of black-hole (BH) mass $M_{BH}$. Moreover, its results remain untested. This paper offers a simple well-tested empirical relation essentially free of assumptions, found using available data for 59 BHs at z>5.7 and the so-called Salpeter relation. We find that accretion rate $\dot{M} \propto M_{BH} (1+z)^3$ consistent with conventional astrophysics that $\dot{M}$ is a direct function of both $M_{BH}$ and ambient gas density. We apply it to the 20 high-z and a subset of Palomar-Green quasars. Comparative tests show that all empirically determined $\dot{M}$ and radiative efficiency $\varepsilon$ pass the tests, but the theoretical determinations fail in most cases. A secondary relation defines Eddington ratio $\lambda$ as a function of z and $\varepsilon$. Consistent with these relations, spline regression analysis of Kozlowski's data for ~132,000 quasars at z<2.4 shows that $\lambda$ and $\varepsilon$ are both functions of $M_{BH}$ and z. The results show that bigger BHs accrete more efficiently than smaller ones. For BHs $>10^9$ solar masses, we get average $\varepsilon$ ~0.23 at z>5.7 and ~0.84 at z<0.005. Notably, the empirical relations predict a mass-inflow rate of 0.11-0.21 sun's mass/year onto the black hole in M87 at z~0.004 that matches its reported Bondi accretion rate determined using observed density and temperature profiles.

# 1. INTRODUCTION

Supermassive black holes (SMBHs) grow by accreting gases and particles, a part of which is radiated away (e.g. Salpeter 1964). The mass-accretion rate ($\dot{M}$) of an active galactic nucleus (AGN), a key factor in understanding the growth of SMBHs, is not measured directly but is inferred from its bolometric luminosity ($L_{bol}$) if its radiative efficiency $\varepsilon$ is known. Soltan (1982) showed that the overall average $\varepsilon$ can be estimated by comparing the integrated AGN luminosity to the total mass accreted by black holes. It is, however, unclear to what extent $\varepsilon$ for individual quasars differs from the canonical value of 0.1 and whether $\varepsilon$ is a function of a black hole's mass, redshift, or spin (e.g. Raimundo et al., 2012). Circumventing these difficulties, several works have attempted to estimate the mass-inflow rate $\dot{M}_{disk}$ for individual quasars using accretion-disc theory and scaling relations between $\dot{M}_{disk}$ and emitted radiation properties such as optical monochromatic luminosities (e.g. Collin et al, 2002; Bian & Zhao, 2002, 2003; Davis & Laor, 2011, Netzer & Trakhtenbrot, 2014, Trakhtenbrot, Volonteri, & Natarajan, 2017). And once a quasar's $\dot{M}_{disk}$ is thus estimated, its radiative efficiency $\varepsilon$ can be inferred from its bolometric luminosity $L_{bol}$.

In particular, Davis and Laor (2011; DL11) developed such a theoretical scaling relation and applied it to 80 Palomar-Green (PG) quasars at z<0.5. Trakhtenbrot, Volonteri, and Natarajan (2017; TVN17) used a slightly modified version to estimate $\dot{M}_{disk}$ and $\varepsilon$ for 20 highest-redshift (z >5.3) quasars. A perplexing characteristic of this relation (Eq.1 in TVN17; Eq.7 in DL11) is that it implies that a quasar's $\dot{M}_{disk}$ is an inverse function of its mass $M_{BH}$ such that the smallest of the 20 quasars is found to accrete at the highest rate (see Table1, TVN17). This result is counter intuitive since accretion rate is a function of a black hole's gravitational reach that increases with its mass. Moreover, the inferred values of $\varepsilon$ for the 20 high-z quasars differ by more than 2 orders of magnitude from 0.003-0.44 that put to question the utility of the widespread use of an average fixed value of 0.1 for $\varepsilon$ in estimating accretion rates from bolometric luminosities. The results remain largely untested, except for the showing that the resultant average $\varepsilon$ for quasars is close to the canonical value. The obvious question is: are these estimates realistic or an artifact of the scaling relation used, and are the wide variations in $\dot{M}_{disk}$ and $\varepsilon$ functions of a quasar's mass, redshift, or some other parameter or simply a result of uncertainties in the determinations of the values of the inputs in the scaling relationships. Raimundo et al. (2012) discuss the considerable uncertainties involved in estimating accretion rates using theoretical scaling relations that are themselves based on numerous assumptions and approximations and using observed properties of quasars that suffer from substantial uncertainties. They concluded that "more observational results are needed to constrain the evolution of black holes and AGN properties".

This paper takes a markedly different approach to estimate accretion rate $\dot{M}$ and $\varepsilon$ for individual quasars. It does not invoke any accretion-disc theory, but offers an empirically derived relation between $\dot{M}$ and quasar's mass $M_{BH}$ and redshift z. This relation is an outcome of a well-tested empirical relation defining $M_{BH}$ as a function of a quasar's age t or redshift z and mass of its seed, found by Aggarwal (2022; A22) using available mass and age data for 59 SMBHs at z >5.7 and the so called Salpeter relation. It is free of assumptions and approximations that could plausibly cloud the validity of the results. A complementary empirical relation is derived expressing the Eddington ratio $\lambda$ as a function of z and $\varepsilon$, from which a quasar's $\varepsilon$ can be derived if its $\lambda$ is known. Having derived these relations, we apply the primary empirical relation to the set of 20 high-z quasars compiled by TVN17 and compare the resulting "empirical" values of Ṁ and ε to their "theoretical" counterparts determined by TVN17. We test the 2 sets of results using several criteria taking into consideration uncertainties in the input data used to derive Ṁ and ε. Then we further test the results and the implications of the empirical relations. The empirical results yield a mean ε ~0.26 for the 20 high-z quasars that is significantly higher than the often used value of 0.1. We validate this result by applying the complementary empirical relation with ε ~0.26 to 59 high-z quasars and comparing the resulting distribution of λ to the observed distribution of λ for 50 high-z quasar in Shen et al. (2019). Also, the empirical relations imply that λ and ε are functions of z and $M_{BH}$, and we find that to be the case by applying spline regression analysis to Kozlowski's (2017) catalog of λ and z values for 132,000 quasars at z<2.4. Thereafter, we apply the empirical relations to a limited set of 80 Palomar-Green (PG) quasars at z~0.35 and compare the results with their theoretical counterparts obtained by DL11. Lastly, we use the empirical relations to determine the mass-inflow rate onto the black hole in Messier 87 for which the Bondi accretion rate has been determined using observations of temperature and density profiles. We end with conclusions and a summary and new insights into the properties and growth of SMBHs.

## 2. EMPIRICAL SCALING RELATIONS

A22 found an empirical relation defining a SMBH's mass $M_{BH}$ as a function of its age t or redshift z and mass $M_S$ and time $t_S$ or redshift $z_S$ of inception its seed, using publicly available mass and age data for 59 high-z (>5.7) SMBHs and the commonly-used Salpeter relation. Of these 59 SMBHs, 36 have common characteristics that helped define 36 equations relating M as a function of t with 3 common parameters. Solving the 36 equations, yielded the following relation.

$$M_{BH} = M_S \exp[14.6(t - t_S)/t] \qquad (1)$$

The above relation can also be expressed as a function of z instead of t (see A22). The best fit to the data produced $t_S$ ~100 million years (Myr) or z ~30. This relationship was extensively tested by applying to SMBHs at z as low as nearly zero

From Eq. 1, we get a SMBH's spot or instantaneous mass-accretion rate $\dot{M} = dM_{BH}/dt$:

$$\dot{M} = (14.6\, t_S/t^2)\, M_{BH} \qquad (2)$$

Where Ṁ is in $M_\odot$/Myr, $t_S$ and t in Myr, and $M_{BH}$ in solar masses. Using the approximation $1/t \propto (1+z)^{3/2}$ for high z (Bergström and Goober, 2006) with the cosmological parameters (Hubble constant =67.4 km/s/Mpc and matter density parameter $\Omega_m$ =0.315) determined by the Planck group (2020) and $t_S$ =100Myr, we get an empirically derived relation defining Ṁ as a function of $M_{BH}$ and z.

$$\dot{M}\,(M_\odot/yr) = 4.96 \times 10^{-12}\, M_{BH}\, (1+z)^3 \qquad (3)$$

Note that Ṁ is in $M_\odot$/yr, $M_{BH}$ in solar masses $M_\odot$. In Eq.3 Ṁ is directly proportional to $M_{BH}$, in contrast to the inverse relationship between $\dot{M}_{disk}$ and $M_{BH}$ implied by the theoretical relation used by DL11 and TVN17; a finding consistent with conventional astrophysics that Ṁ increases with $M_{BH}$ as a BH's gravitational reach increases. Furthermore, as per the Standard Cosmological Model, the density of ordinary matter at any z relative to that at the current epoch scales as $(1 + z)^3$. Thus, Eq.3 indicates that $\dot{M}$ is a function of the ambient gas density and decreases with z in proportion to the decrease in density. Note that Eq.2 and 3 yield slightly different values of $\dot{M}$ because of the approximation used to convert t into z.

Once a BH's accretion rate $\dot{M}$ is determined using Eq.3, its $\varepsilon$ can be inferred from its bolometric luminosity $L_{bol}$. Radiative efficiency $\varepsilon$ is conventionally defined with respect to the mass inflow rate, and a BH's accretion rate $\dot{M}$ is smaller by a factor of $(1 – \varepsilon)$. Hence, $\varepsilon/(1- \varepsilon) = L_{bol} / (\dot{M}c^2)$ where c is the velocity of light (see Shankar et al, 2010). Substituting the value of c, and expressing $L_{bol}$ in units of $10^{46}$ ergs/s and $\dot{M}$ in $M_\odot$/yr, one gets:

$$\varepsilon / (1- \varepsilon) = 0.176\, L_{bol} / \dot{M} \quad (4)$$

Furthermore, we can define Eddington ratio $\lambda = L_{bol} / L_{Edd} = \dot{M}/\dot{M}_{Edd}$ as a function of z and $\varepsilon$ using Eq.3 and 4 and Eddington luminosity $L_{Edd} = 1.3 \times 10^{38}\, M_{BH}$ in ergs/s.

$$\lambda = 2.18 \times 10^{-3} (1 + z)^3\, \varepsilon /(1- \varepsilon) \quad (5)$$

Conversely, we can use Eq. 5 to estimate radiative efficiency $\varepsilon$ from a quasar's Eddington ratio $\lambda$. Note that Eq.4 is a generic equation, whereas Eq.5 is derived empirically based on Eq.3.

## 3. INPUT DATA AND DERIVED PROPERTIES

Table 1 in TVN17 lists the input data ($M_{BH}$ and optical luminosities), the derived mass-inflow rate for 20 quasars at z>5.3. Note that the symbol for radiative efficiency used here is $\varepsilon$ and $\eta$ in TVN17. Multiplying their $\dot{M}_{disk}$ by $(1- \varepsilon)$ gives a BH's theoretical accretion rate $\dot{M}$. We used their $M_{BH}$ and z data to calculate the empirical $\dot{M}$ for each one of the 20 quasars using Eq.3. Having thus obtained $\dot{M}$, we calculated $\varepsilon$ for each of the 20 quasars using Eq.4 and $L_{bol}$ determined by TVN17. For brevity, a derived property using the empirical relation will henceforth be referred to as the empirically derived property and that resulting from the theoretical relation as the theoretically derived property. The input $M_{BH}$ and z data from TVN17 and the resulting empirically derived $\dot{M}$ and $\varepsilon$ and the corresponding theoretical values (in parentheses) are given in Table 1 (this study). In addition, Table 1 lists for each BH its Eddington ratio $\lambda$ calculated using the $L_{bol}$ from TVN17 and $L_{Edd} = 1.3 \times 10^{38}$ $M_{BH}$ in ergs/s with $M_{BH}$ in solar masses. The BH ages were determined using the cosmological parameters of the Planck group (2020). Lastly, Table 1 also shows the empirical and theoretical (in parentheses) accretion times $t_{acc} = M_{BH} / \dot{M}$ for each BH. Note that for convenience the BH name in column 1 is preceded by a number corresponding to the row in which it is listed.

The masses of the 20 BHs in Table 1 were determined by TVN17 using a prescription by Trakhtenbrot and Netzer (2012) based on data for low-z BHs. It is instructive to compare the TVN17 determinations of $M_{BH}$ to other sources. A22 lists the $M_{BH}$ and z data for 19 of the 20 BHs compiled from alternate sources cited in Table 1 in A22. Each of these 19 BHs is identified by an alternate BH # in column 2 of Table 1 (this study) that corresponds to the BH number in Table 1 of A22. Wang et al. (2015) is an alternate source for the remaining BH identified by W. A comparison of the 2 sets of $M_{BH}$ and z data reveals that the redshifts are identical or nearly identical, but the $M_{BH}$ computed by TVN17 are systematically higher on the average by a factor of ~1.6 except for BH#13 for which it is smaller by a factor of ~3. Hence, if the alternate sources of $M_{BH}$ were used, the empirical $\dot{M}$ for the 19 BHs would be lower in general by a factor of ~1.6 and the theoretical $\dot{M}$ higher by a similar factor; and the empirical $\varepsilon$ would be ~40% higher and the theoretical $\varepsilon$ lower by a similar percentage than those in Table 1. The BH#13 is the smallest of the 20 in Table 1. As determined by TVN17, its $\dot{M}_{disk}$ is the highest and its $\varepsilon$ lower by an order of magnitude than the rest of the BHs. Using its alternative $M_{BH}$ ($3\pm0.1 \times 10^8 M_\odot$) reduces its theoretical $\dot{M}$ by ~2/3 and increases its $\varepsilon$ by a factor of ~3, bringing them within the range for the other 19 BHs. Hence, to minimize the skewing of the results by this one BH we used its alternative $M_{BH}$ to estimate its theoretical and empirical $\dot{M}$, $\varepsilon$, and $\lambda$. Note that in the comparative analyses that follow this change in the $M_{BH}$ of BH#13 favors the theoretical results.

## 4. COMPARATIVE TESTS OF DERIVED PROPERTIES

### 4.1 Mass Accretion Rates

In several cases the empirical and theoretical $\dot{M}$ are comparable, but in most cases the theoretical $\dot{M}$ are significantly higher and in several cases by an order of magnitude. The differences between the two sets of $\dot{M}$ would be even more pronounced if the alternate measures of $M_{BH}$ were used. One way to test whether the $\dot{M}$ values are reasonable and meet the proverbial "smell test" is to assess the accretion times ($t_{acc} = M_{BH}/\dot{M}$) implied by the two sets of $\dot{M}$. Figure 1 shows the distribution of the theoretically and empirically derived $t_{acc}$ relative to BH age. The theoretical $t_{acc}$ vary widely by more than 2 orders of magnitude and are internally inconsistent. Seven BHs (#6, 13, 14, 15, 17, 18, 20) have very short $t_{acc}$ (~16-62 Myr); four (#7, 8, 10, 16) have somewhat longer (~121-141 Myr); one (#12) exceeds its age, while some have $t_{acc}$ shorter than but comparable to their ages. In particular, of the 20 BHs, 11 have $t_{acc}$ <150 Myr and <100 Myr if the alternate $M_{BH}$ were used to estimate their theoretical $\dot{M}$. Such extremely short $t_{acc}$ are problematic. The seeds of SMBHs are generally thought to have formed at z <20 when the universe was < 200 Myr old and recent growth models (e.g. Pacucci and Loeb, 2020; Zubovas and King, 2021) assume and empirical evidence (A22) suggests that the seeds formed at or near z=30 when the universe was only ~100 Myr old. The average age (t) of these 11 BHs is ~910 Myr. Hence, the seeds of these BHs potentially had >700 Myr to grow. It is difficult, if not impossible, to reconcile the derived theoretical $t_{acc}$ <150 Myr with an accretion time of >700 Myr potentially available to these quasars unless their seeds remained essentially dormant for long periods or formed shortly before the redshift at which the BHs are observed. Neither possibility is likely in view of the fact that BH#1 that is younger by >150 Myr is found to be actively accreting and some of the BHs have been accreting for the bulk of the time potentially available to them.

Table 1
Comparison of BH properties derived using empirical and theoretical scaling relationships

| BH Number -Name | BH # in A22 | Log $M_{BH}$ ($M_\odot$) | Red shift z | Edd. ratio $\lambda$ | Accretion rate $M_\odot$/yr $\dot{M}$ | Radiative efficiency $\varepsilon$ | Accretion Time Myr $t_{acc}$ | Age Myr t |
|---|---|---|---|---|---|---|---|---|
| 1- J1120+0641 | 4 | 9.58 | 7.097 | 0.19 | 9.84 (9.76) | 0.13 (0.144) | 386 (390) | 745 |
| 2- J1148+5251 | 26 | 9.93 | 6.407 | 0.16 | 16.93 (13.22) | 0.15 (0.189) | 503 (644) | 851 |
| 3- J0100+2802 | 31 | 10.34 | 6.3 | 0.15 | 42.21 (47.01) | 0.15 (0.139) | 518 (465) | 870 |
| 4- J0306+1853 | W | 10.32 | 5.363 | 0.15 | 26.69 (48.42) | 0.21 (0.126) | 783 (431) | 1069 |
| 5- J0050+3445 | 32 | 9.61 | 6.253 | 0.21 | 7.61 (12.69) | 0.20 (0.131) | 535 (321) | 879 |
| 6- J0836+0054 | 58 | 9.67 | 5.81 | 0.37 | 7.2 (120.4) | 0.36 (0.032) | 650 (39) | 964 |
| 7- J0353+0104 | 47 | 9.38 | 6.072 | 0.33 | 4.22 (16.95) | 0.26 (0.079) | 568 (141) | 913 |
| 8- J0842+1218 | 48 | 9.47 | 6.069 | 0.24 | 5.17 (23.00) | 0.22 (0.065) | 570 (128) | 913 |
| 9- J2348−3054 | 8 | 9.46 | 6.886 | 0.084 | 7.02 (6.14) | 0.07 (0.083) | 411 (470) | 775 |
| 10- J0305−3150 | 15 | 9.15 | 6.605 | 0.30 | 3.08 (11.63) | 0.23 (0.077) | 487 (121) | 819 |
| 11- P036+03 | 18 | 9.50 | 6.527 | 0.35 | 6.69 (10.20) | 0.29 (0.197) | 478 (310) | 831 |
| 12- P338+29 | 14 | 9.72 | 6.658 | 0.06 | 11.33 (2.88) | 0.06 (0.199) | 463 (1820) | 810 |
| 13- J0005−0006 | 57 | 8.48 | 5.844 | 0.78 | 0.48 (62) | 0.76 (0.01) | 582 (5) | 957 |
| 14- J1411+1217 | 56 | 9.24 | 5.903 | 0.51 | 2.83 (62.79) | 0.42 (0.031) | 614 (28) | 941 |
| 15- J1306+0356 | 52 | 9.20 | 6.017 | 0.34 | 2.72 (32.61) | 0.31 (0.036) | 583 (49) | 923 |
| 16- J1630+4012 | 49 | 9.20 | 6.058 | 0.28 | 2.76 (12.30) | 0.30 (0.075) | 574 (129) | 915 |
| 17- J0303−0019 | 46 | 8.72 | 6.079 | 0.47 | 0.92 (32.73) | 0.38 (0.017) | 570 (16) | 911 |
| 18- J1623+3112 | 36 | 9.32 | 6.211 | 0.27 | 3.88 (34)) | 0.24 (0.036) | 538 (62) | 886 |
| 19- J1048+4637 | 37 | 9.83 | 6.198 | 0.44 | 12.51 (8.45) | 0.35 (0.444) | 540 (800) | 889 |
| 20- J1030+0524 | 30 | 9.28 | 6.302 | 0.29 | 3.68 (33.93) | 0.26 (0.036) | 518 (56) | 870 |

In column 1, BH name is preceded by a corresponding row number for easier identification. In column 2, BH # corresponds to the # for the same BH in Table 1 of A22, except that identified by W= Wang et al. (2015). Theoretically derived properties are shown in parentheses, and corresponding empirical not in parentheses.

BH mass $M_{BH}$, redshift z, theoretical $\varepsilon$, and theoretical $\dot{M}$ = (1- $\varepsilon$) $\dot{M}$disk from TVN17

Empirical $\dot{M}$ derived using Eq.3, and empirical $\varepsilon$ derived using Eq.4.and $L_{bol}$ from TVN17.

Eddington ratio $\lambda = L_{bol}/L_{Edd}$ calculated using $L_{Edd} = 1.3 \times 10^{38} M_{BH}$ and $M_{BH}$ in solar masses.

Accretion time scale $t_{acc} = M_{BH}/\dot{M}$.

*Note: TVN17 estimates of $M_{BH}$ are systematically higher on the average by factor of ~1.6 than those in A22, except BH in row #13 for which TVN17 estimate is lower by a factor of ~3. See text for the effect on $\dot{M}$ and $\varepsilon$ values if alternate estimates of $M_{BH}$ were used.*

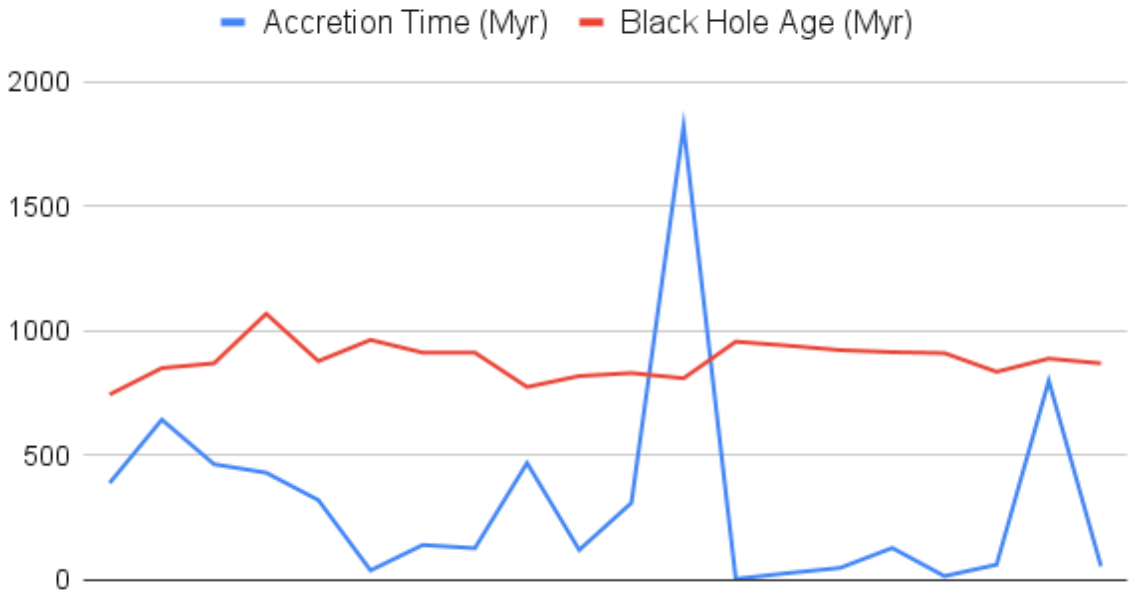

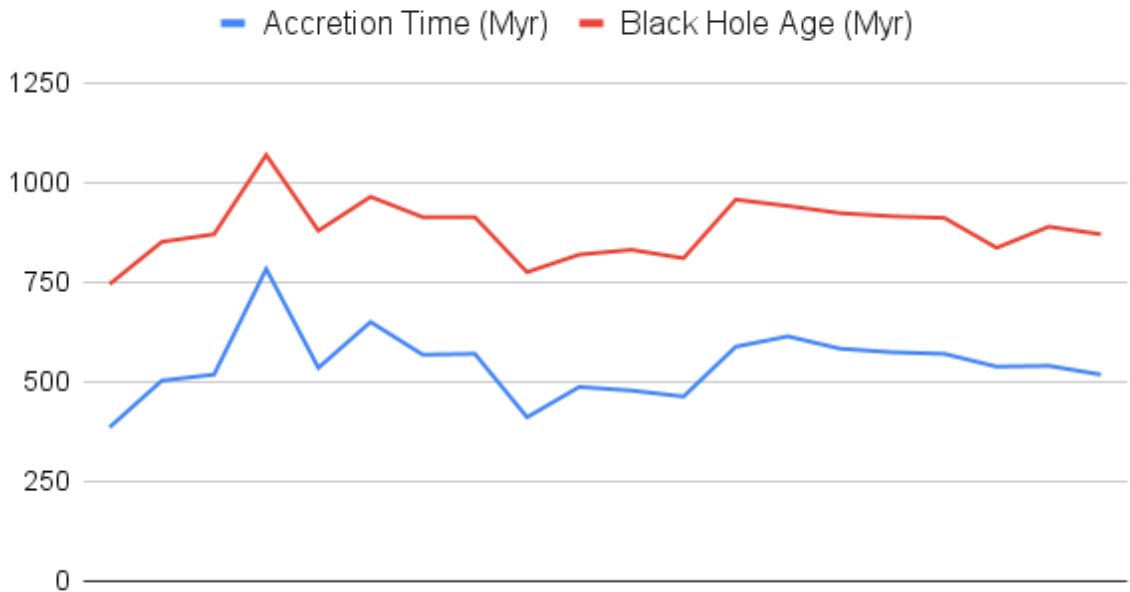

Fig.1: Theoretically and empirically determined accretion time $t_{acc}$ of 20 high z (>5.3) black holes (BH) relative to their ages. The theoretical $t_{acc}$ bear no correlation with BH ages and are internally inconsistent: one exceeds its age; 11 others have accretion times that are only ~2-16% of their life spans; while some have $t_{acc}$ comparable to their ages. In contrast, the empirical $t_{acc}$ show a close correlation with BH ages and are systematically lower but within a factor of <~2 of their ages.

In contrast, Fig.1 shows a close correlation between empirical $t_{acc}$ and BH ages. The youngest (#1) yields the shortest $t_{acc}$ ~386 Myr, and the oldest (#4) has the longest $t_{acc}$ ~783Myr. And for the 11 BHs in question for which the theoretical $t_{acc}$<150 Myr, the empirical $t_{acc}$ range from ~487- 650 Myr that are shorter but comparable to the growth times potentially available to them. We conclude that no inconsistencies or contradictions were found in the empirically determined $t_{acc}$. And since empirical $t_{acc}$ is independent of BH mass (see definition of $t_{acc}$ and Eq.3), the alternate $M_{BH}$ would yield exactly the same empirical $t_{acc}$ as those in Table1.

It is instructive to compare the empirical spot value $\dot{M}$ for each BH with its accretion rate $<\dot{M}>$ averaged over its lifespan. Since the seeds of SMBHs are expected to be smaller by orders of magnitude than their $M_{BH}$ at the redshifts at which they are observed, the average accretion rate $<\dot{M}>$ is simply $<\dot{M}> = M_{BH} / (t-t_s)$, where t is the BH's age and $t_s$ the origin time of its seed. Adopting $t_s$ =100Myr, it turns out that spot empirical $\dot{M}$ for each BH is systematically larger than its $<\dot{M}>$ by a factor of $\leq 1.5$. This comparison suggests that BH growth is apparently not episodic, but the accretion rate increases with time from the inception of the seed to the age at which these BHs are observed. In fact, we can trace the growth of a BH seed from its inception using the following Eq.6 in which $\dot{M}$ is expressed as a function of a BH's redshift z and mass $M_S$ and redshift $z_S$ of its seed by substituting Eq.1 into Eq.3 and converting t and $t_S$ into z and $z_S$.

$$\dot{M} (M_\odot/yr) = M_S \, 4.96 \times 10^{-12} (1 + z)^3 \, \text{Exp} \, 14.6 \, [1 - (1 + z)^{3/2} / (1+ z_S)^{3/2}] \qquad (6)$$

Figure 2 shows the accretion rate $\dot{M}$ of a unit (solar) seed mass as a function of z starting at z=30. The results show that $\dot{M}$ initially increases exponentially, reaches a broad peak between z~8.5 to 6 and steadily decreases thereafter. The nominal peak occurs at z~7.25 with $\dot{M}$=8.22x10$^{-4}$ $M_\odot$/yr per unit (solar) seed mass. Hence, $t_{acc}$ calculated using spot values of $\dot{M}$ for quasars at z > ~6 should be shorter than their life spans as found above, and $t_{acc}$ for BHs at very low z (e.g. z <1) should be longer than their life spans as found later.

## 4.2 Radiative Efficiencies

Next, a comparison of the two sets of determinations of radiative efficiencies shows that in several cases the two determinations are comparable, but in most cases the empirical values are significantly higher than their theoretical counterparts. The average empirical $\epsilon$ is ~0.26, and the theoretical ~0.11. The theoretical average is almost identical to the often used canonical value of 01, a result used by TVN17 to support their overall results and methodology. Shankar et al. (2010), however, investigated the characteristic radiative efficiency $\epsilon$, Eddington ratio λ, and duty cycle of high-z AGNs and concluded that to simultaneously reproduce the observed luminosity function and bias, $f / \lambda \gtrsim 0.7$ where $f = \epsilon/(1 - \epsilon)$. They further affirmed that these findings are robust against "uncertainties in the obscured fraction of AGNs or in the precise value of the mean bolometric correction". The average λ for the 20 BHs is 0.297. Hence, the average $\epsilon$ should be > 0.17. The theoretical average of $\epsilon$ is clearly well below this lower limit and the empirical average well above it. Applying this test individually to each of the 20 BHs shows that all of the empirically determined $\epsilon$ (including the singled out BH #13) meet this criteria, whereas the theoretical estimates of $\epsilon$ do not in 11 of the 20 cases And if the alternate measures of $M_{BH}$ were used instead, only 3 of the 20 theoretical estimates of $\epsilon$ would survive the test; whereas all empirical estimates would still pass the test.

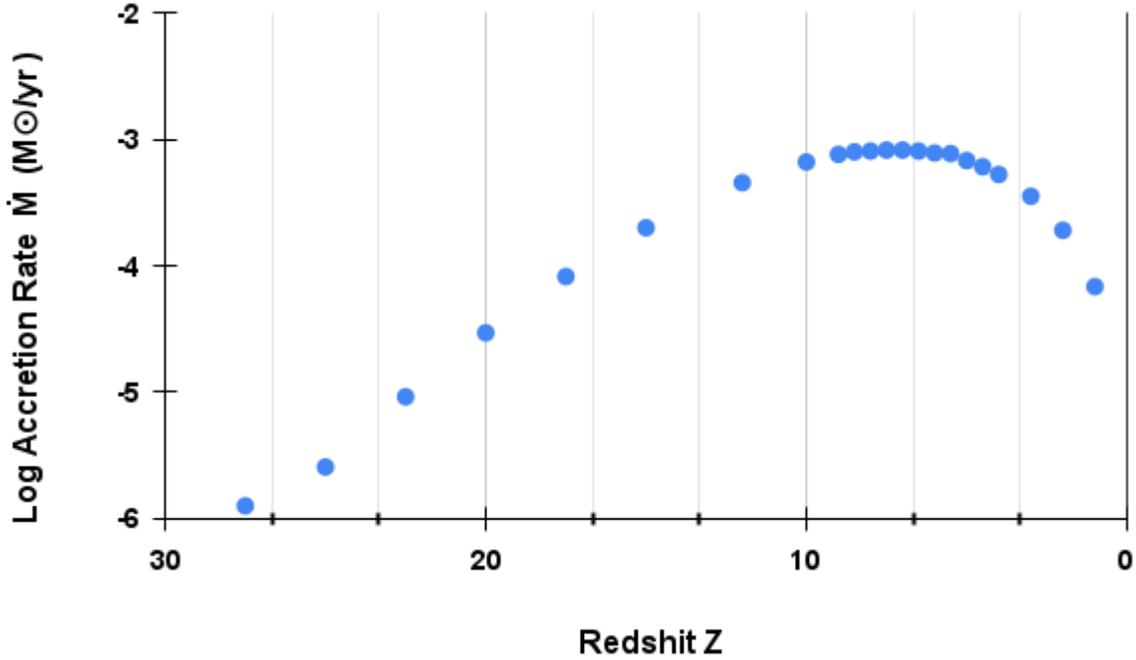

Fig.2: Log of spot accretion rate Ṁ for a BH seed of unit (solar) mass M☉ as a function of redshift starting at z=30 based on Eq.6. Initially Ṁ increases exponentially, reaches a broad peak between z ~8.5 and 6, and steadily decreases thereafter towards z=0.

To further assess the validity of Eq.3 and the empirical mean value of ε ~0.26 for high-z quasars, we applied Eq.5 with ε=0.26 to 59 high-z BHs complied by A22 and compared the resulting distribution of λ to the observed distribution of λ for 50 high-z BHs determined by Shen et al. (2019).. Table 1 in A22 lists the properties of the target sample of 59 BHs with references therein for the sources of data. Their redshifts range from ~5.784 to ~7.64 and include the 3 quasars at the highest z (>7.5) observed thus far. The Shen et al. (2019) sample has a narrower z-range of ~5.63-6.42 and includes some of the quasars in the target sample. Note that a BH's empirical λ thus calculated is solely a function of its redshift (Eq.5) and does not take into account variations in a BH's real ε from the mean value used. On the other hand, a BH's observed λ depends upon uncertainties in the monochromatic luminosity used, the correction factor applied, and BH mass $M_{BH}$. The empirical λ for the 59 quasars thus determined range from ~0.24-0.5 with a mean value of ~0.31. In comparison, the distribution of "observed" values of λ for 50 quasars shown in Fig.9 of Shen et al. (2019) define a broad peak with a median value of ~0.32 and >80% of λ between ~0.16 and ~0.63. In fact, all but 2 of the observed values of λ could be within this range taking into consideration the 1σ uncertainty in λ shown in Table 3 of Shen et al. (2019). The empirical and observed median values of λ are almost identical, and the broad agreement between the two distributions is surprisingly good given the uncertainties in the determination of observed λ and the fact that a fixed mean value of ϵ is used to calculate empirical λ. This agreement re-enforces the result that the median value ϵ of high-z SMBHs is close to the empirically determined value of ~0.26 rather than the often assumed canonical value of 0.1.

## 4.3 Eddington ratio and radiative efficiency as functions of z and $M_{BH}$

Equation 5 implies that λ decreases as a BH ages or its redshift decreases. Whether this is indeed the case has remained an open question as exemplified by the following case studies. Shen et al (2019) did not find a significant difference in the distribution of λ for their high-z (>5.65) sample of 50 quasars and a luminosity-matched control sample of quasars at z= 1.5-2.3; whereas Willott et al, (2010) found a marked difference in the distribution of λ at z=2 and z=6, with the latter quasars accreting in general at higher λ than the former. Furthermore, Shen et al. (2008) obtained mean values of λ ranging from ~0.079-0.25 with typical widths of < 0.3dex for AGN's in different luminosity bins at z ~4.5-0.1. In contrast, Kelly et al. (2010) found that compared to previous work their inferred λ for a sample of 9886 quasars at 1< z < 4.5 corrected for completeness is shifted towards lower values of λ peaking at ~0.05 with a dispersion of ~0.4 dex. In fact, Schulze and Wisotzki [56] found a z-dependence of λ in the Shen et al. (2008) data, noting that restricting their sample to z ≤ 0.3 gives a lower mean λ of ~0.067 with a dispersion of ~0.43dex. Furthermore, Trakhtenbrot and Netzer (2012) found a steep rise in λ with z up to z ≃ 1 with smaller BHs accreting at higher values of λ. Perhaps the strongest statistical evidence thus far in favor of a probable decrease in λ with z is the observation that the λ for high-z quasars are almost all ≥0.1 (e.g. Shen et al., 2019), whereas λ for AGNs at 1< z <2 determined by Suh et al. (2015) cover a much wider range down to 0.001.

Ascertaining whether λ indeed decreases with z has been problematic for several reasons. First, multiple uncertainties affect the determination of λ resulting from uncertainty in the monochromatic luminosity used and the bolometric correction factor applied. Second it is difficult to cross-compare the results of two studies that use different bolometric correction factors. And since Eddington luminosity is a function of $M_{BH}$, λ could also be a function of $M_{BH}$ that may obscure the dependence of λ on z as found by Trakhtenbrot and Netzer (2012) for BHs at z ≤ 1. The available λ data for AGNs at z > 3 are not large enough to separate the BHs into narrow $M_{BH}$ bins and meaningfully investigate the dependence of λ on z. Kozlowski's (2017) catalog of properties of 280,000 AGNs in the Sloan Digital Sky Survey, however, provides a subset with generally well constrained λ to perform such an analysis. The subset consists of ~132,000 AGNs at ~ 0.1< z <2.4 with $M_{BH}$ determined using the more reliable Mg II lines and $L_{bol}$ based on a weighted mean of bolometric luminosities from 2 or more monochromatic luminosities. The reported uncertainties in $L_{bol}$ thus determined are generally small and so is the dispersion in λ values except for AGNs at z < 0.5. Figure 3 shows a spline regression plot of λ as a function of z using ggplot2 by Wickham (2016) for BHs in 3 bins of $M_{BH}$ covering a time span of almost 10 billion years. Each line represents the mean value of λ as a function of z, and the grey shaded area shows the 95% confidence interval. The probability that λ is within the shaded area depends upon the number of data points within Δz and the number of BHs in each group. The $1-3 \times 10^8 \, M_\odot$ group (red) has 36,871 AGNs; the $1-3 \times 10^9 \, M_\odot$ green group has 28,799 AGNs; and the $1-3 \times 10^{10} \, M_\odot$ blue group has 523 AGNs. The number of data points or BHs decreases at lower redshifts. The red line or the $1-3 \times 10^8 \, M_\odot$ group shows a clear decrease in λ by more than a factor of 3 from z ~ 2.3 to z <0.5. Notwithstanding the larger dispersion in λ at z<0.5 for the $1-3 \times 10^9 \, M_\odot$ group, the green line also shows a clear decrease in λ from z~2.3 to z < 0.6 by a similar factor. The blue line for the $1-3 \times 10^{10} \, M_\odot$ group is, however, ambiguous given the much larger dispersion in λ for this group. This group has only 523 AGNs compared to tens of thousands in the other two. Nonetheless, the λ of these AGNs are generally smaller by ~ an order of magnitude than the average λ of ~0.30 for the 20 high-z AGNs in Table 1. A fourth $M_{BH}$ group of $1-3 \times 10^7 \, M_\odot$ with 674 BHs (not plotted because of scale differences) in fact shows a more pronounced decrease in λ with z. Thus, the decrease in λ with z predicted by the empirically derived Eq.5 based on data for high-z BHs is apparently borne out by Kozlowski's data for low-z quasars.

Notably, the spline regression in Fig.3 suggests that λ for BHs in the green and blue bins essentially reach a constant at z<0.5. Equation 1 predicts that a BH's $M_{BH}$ increases by only ~7% from z =0.5 to z=0, a time span of ~5 billion years. Hence, its $L_{EDD}$ changes little and so must its $L_{bol}$ for λ to remain essentially constant. On the other hand, Eq.3 predicts that $\dot{M}$ decreases by a factor of ~3 in this time span despite the small increase in $M_{BH}$. Therefore, as per Eq.4 any decrease in $\dot{M}$ must be accompanied by roughly a similar increase in ε/(1- ε) for $L_{bol}$ to remain essentially unchanged. Since $\dot{M}$ is a direct function of z (Eq.3), the implication is that ε should be an inverse function of z. Such a conclusion can also be drawn from Eq.5 with λ constant.

Figure 3 indicates that besides z, λ is apparently also a function of $M_{BH}$. This dependence of λ on $M_{BH}$ is unambiguous at z >1. For example, at z=1.5 the λ value for the red group (smallest BHs) is ~ 3.5 times higher than that for the blue group (largest BHs). Thus, observational data indicates that in addition to being a direct function of z, λ is apparently an inverse function of BH mass. In Eq.5, however, λ is a direct function of z and radiative efficiency ε. Hence, ε should be an inverse function of $M_{BH}$ if Eq.5 is valid. We can test this implication using high-z data for a sufficiently wide range of $M_{BH}$. All but 4 of the 20 BHs in Table 1 have masses within a narrow range of $10^9$-$10^{10}$ $M_\odot$. Hence, ε for 13 additional high-z BHs with $M_{BH}$ $10^8$-$10^9$ $M_\odot$ and for one available with $M_{BH}$ >$10^{10}$ $M_\odot$ were determined using Eq.3 and 4. The 14 additional BHs are listed in Table1 of A22 as BH # 2,5,10,20,21,22,25,28,35,47, 51,54,55 and 57 with references therein for the sources of pertinent data. Figure 4 shows log ε as a function of log $M_{BH}$ for the 34 high-z quasars, 20 of which are from Table1. There is considerable dispersion in ε for a given $M_{BH}$ probably resulting from uncertainties in $L_{bol}$ and $M_{BH}$, the input data used to determine ε. Nonetheless, the observed trend is clear and supports the implication that ε decreases as $M_{BH}$ increases. Thus, the implication that ε is apparently an inverse function of $M_{BH}$ deduced from observational data for low-z quasars and Eq.5 is borne out by the application of Eq.3 and 4 to the data for high-z BHs. A least-squares fit to the data in Fig. 4 yields the following empirical relationship applicable only to high-z BHs at z >5.7 and $M_{BH}$ >$10^8$ $M_\odot$.

Log ε = (2.552 $\pm$ 0.55) - (0.334 $\pm$ 0.06) Log $M_{BH}$                                 (7)

Equation 7 yields ε =0.76 for the smallest and ε =0.13 for the largest BH at z>5.7 in Table 1, in agreement with the empirical ε values in Table 1. The standard deviations in estimating ε for individual BHs using Eq.7 are large reflecting the large dispersion in ε values of BHs with similar masses in Fig.4. Moreover, the following observations support the implication/conclusion derived earlier that ε is apparently also a function of z that Eq.7 does not take into account. Applying Eq.5 to 541 BHs in Kozlowski's catalog (2017) with $M_{BH}$ ≥$10^{10}$ $M_\odot$ at an average redshift z=1.906 and a mean λ=0.0392, we get a mean ε= 0.422 that is significantly higher than ε <0.16 predicted by Eq.6 for similar-size high-z (>5.7) BHs. Furthermore, using the mean λ values for BHs at z=1.5 in each of the 3 bins in Fig.3 and applying Eq.5, we get ε=0.838, 0.736, and 0.595 respectively for BHs in the red, green and blue bins. In comparison, Eq.6 yields ε=0.62, 0.29, and 0.13 for high z (>5.7) BHs respectively in the 3 mass bins. In each case, the BHs at z=1.5 have a significantly higher ε than the high-z BHs, indicating that ε is apparently an inverse function of z in addition to being an inverse function of BH mass.

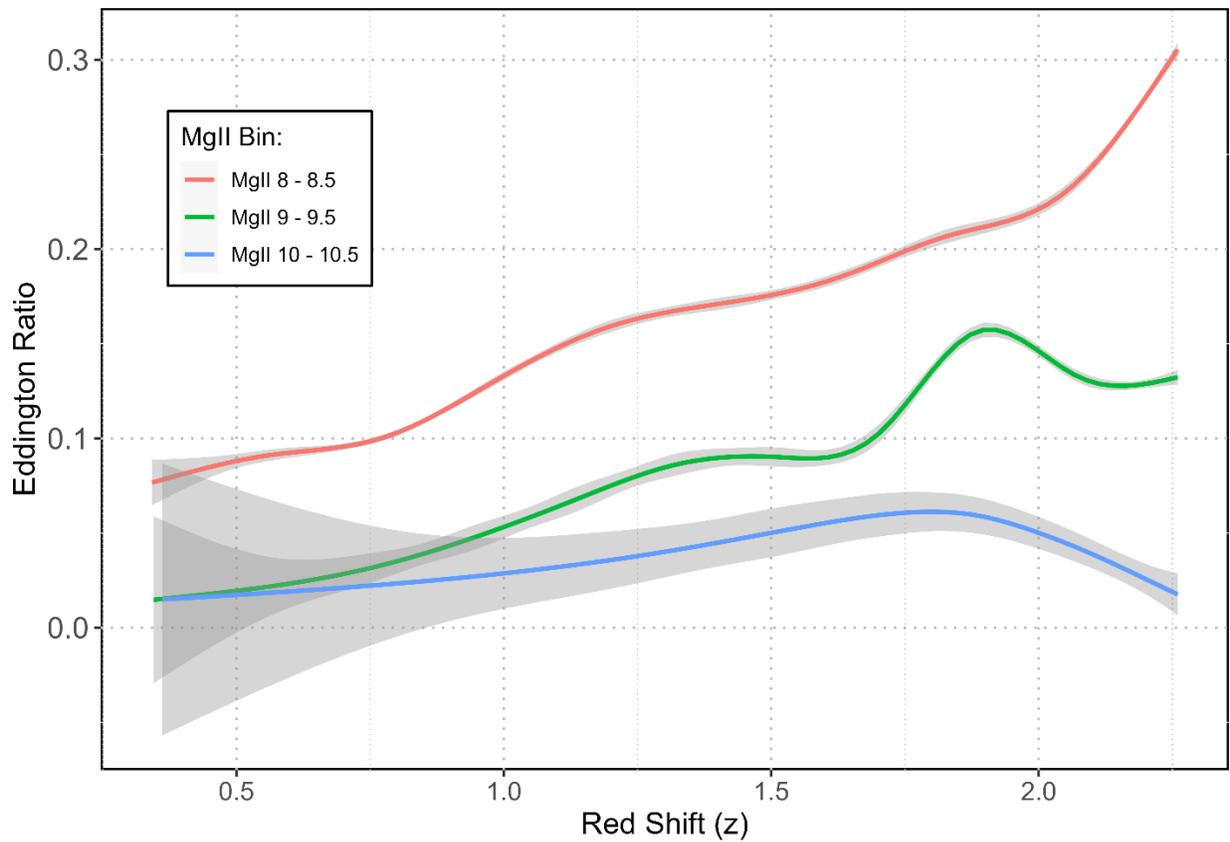

Fig.3: Spline regression plot of Eddington ratio λ versus z using ggplot2 by Wickham (2016). The data are from Kozlowski's (2017) catalog separated into 3 $M_{BH}$ bins. The 1-3x10$^8$ M☉ red group has 36,871; the 1-3x10$^9$ M☉ green group has 28,799; and the 1-3x10$^{10}$ M☉ blue group has 523 AGNs. Each line gives the median value of λ as a function of z, and the grey area around it shows the 95% confidence interval. A fourth $M_{BH}$ group of 1-3x10$^7$ M☉ with 674 BHs (not plotted because of scale differences) in fact shows a more pronounced decrease in λ with z. There are relatively few data points for the blue bin at z < 1. In contrast, the green bin at z ~0.4 has 35 data points.

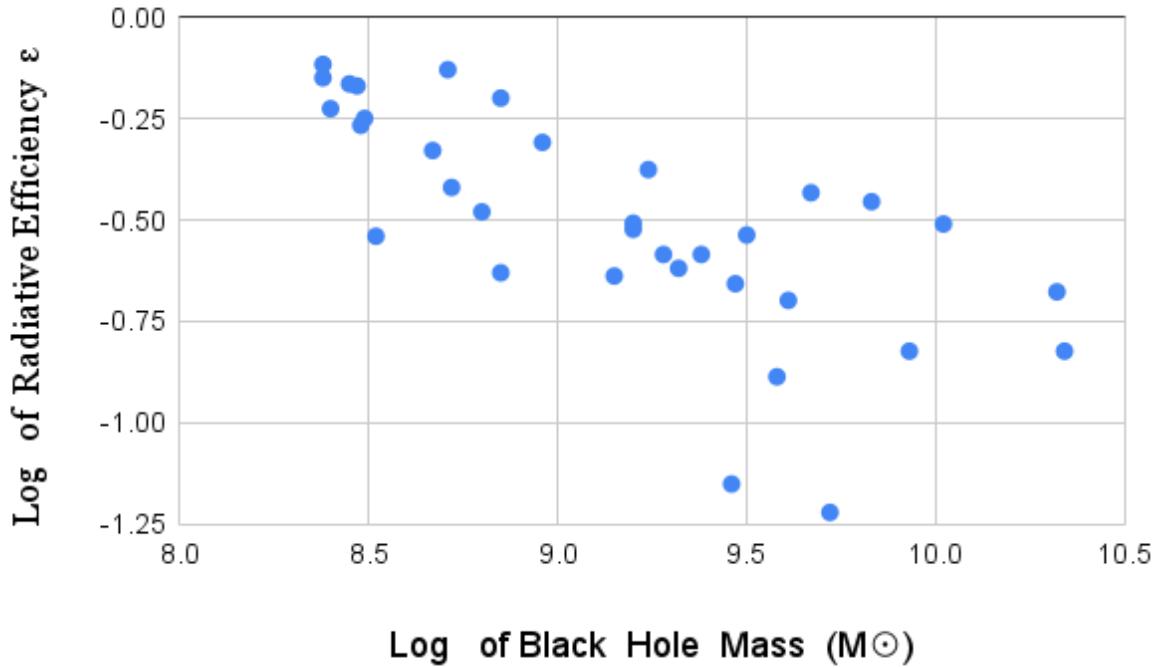

Fig. 4: Log of radiative efficiency ε plotted against log of BH mass $M_{BH}$ for 34 quasars at z >5.3 and mass >$10^8\,M_\odot$. Radiative efficiencies are calculated using Eq. 4 and 3. See text for sources of input data for the 34 BHs. Equation 7 (text) gives a least-squares fit to the data applicable only to BHs at z>5.7 and $M_{BH}$ >$10^8\,M_\odot$.

## 5. COMPARISON OF ACRETION RATES OF PG QUASARS

DL11 derived mass inflow rates $\dot{M}_{disk}$ and radiative efficiencies ε for 80 Palomar-Green (PG) quasars using their (Eq.7) theoretical scaling relation. An exhaustive comparative analysis of their results with those using Eq.3 and 4 (this study) is probably not very useful. The following limited comparison, however, provides a measure of the differences between the two sets of results. Table 1 in DL11 lists 11 PG quasars >$10^9\,M_\odot$ within a narrow $M_{BH}$ and z range; averaging ~$1.64 \times 10^9\,M_\odot$ for $M_{BH}$ and ~0.35 for z. The derived $\dot{M}_{disk}$ range from 1-15 $M_\odot$/yr, averaging ~3.44 $M_\odot$/yr; the radiative efficiencies range from ~0.21-0.54 yielding an average ε=0.35; and the bolometric luminosities $L_{bol}$ range from ~$10^{46}$ –$1.3 \times 10^{47}$ with an average of ~ $5.36 \times 10^{46}$ ergs/s. Multiplying the average $\dot{M}_{disk}$ by (1- ε), we get an average theoretical accretion rate $\dot{M}$ = 2.23 $M_\odot$/yr. In contrast Eq.3 yields an average empirical $\dot{M}$ ~0.02 $M_\odot$/yr, which is 2 orders of magnitude lower. Incidentally, the average theoretical $\dot{M}_{disk}$ for the 11PG quasars is also ~2 orders of magnitude greater than the Bondi accretion rate for the BH in Messier 87 that has a mass somewhat larger than the 11 PG quasars albeit at a lower redshift; whereas its empirically determined mass-inflow rate matches the Bondi rate as discussed and shown later. Furthermore, the theoretically determined $\dot{M}$ yields an accretion time scale $t_{acc}$ < a billion years, whereas the average age of the 11 quasars is almost 10 billion years. The two time scales are not reconcilable, unless the seeds remained dormant for much of the life spans of the quasars or accreted at rates much smaller than the spot value. Such a possibility is untenable in light of the result that similar-size high-z quasars (Table 1) were accreting at rates comparable or higher than that for the 11 PG quasars. On the other hand, the empirically derived $\dot{M}$ yields $t_{acc}$ exceeding the life-span of the universe as expected and pointed out earlier. Figure 2 shows that these quasars were accreting at higher rates than their spot $\dot{M}$ during most of their life spans Hence, the accretion rate $<\dot{M}>$ averaged over the life spans of these quasars should be much higher that their spot accretion rate $\dot{M}$; which indeed is the case with $<\dot{M}>$ ~0.2 M☉/yr or an order of magnitude higher than the spot value.

Furthermore, the average Eddington ratio λ for the 11 PG quasars in DL11 is ~0.25, which is ~ an order of magnitude higher than the average λ for similar-size quasars at similar redshifts in Fig.3 (see green bin at z~0.4). Kozlowski (2017) in fact lists 35 quasars in the 1-3x$10^9$ M☉ green bin within a narrow z range of ~ 0.492-0.345 having a mean λ value of ~0.0118 with 80% < 0.012. This difference in λ values apparently results from the fact that the $L_{bol}$ determined by DL11 for the 11 quasars are on the average higher by more than an order of magnitude than the average $L_{bol}$ for the 35 quasars in Kozlowski's catalog. Hence, if Kozlowski's average $L_{bol}$ value were to be used for the 11 PG quasars, the resulting average theoretical ε would be more than an order of magnitude smaller than the average value of ~0.35 determined by DL11. In contrast, Eq.5 (this study) yields an average ε ~0.66 for quasars in the green bin in Fig.3 at z=0.4 using λ=0.0118. Lastly, DL11 found that ε apparently increases with BH mass $M_{BH}$, whereas the results of this study show just the opposite. Raimundo et al. (2012) suggested that this finding was most likely an artefact of the parameter space covered by their PG quasar sample. We submit that it is an artifact of the theoretical scaling relation in which $\dot{M}_{disk}$ is an inverse function of $M_{BH}$ such that of 2 BHs with similar optical luminosities, the $\dot{M}_{disk}$ for the smaller BH is larger than that for the bigger BH. As a result, ε derived from their bolometric luminosity is higher for the bigger BH than that for the smaller BH.

## 6. ACCRETION RATE OF BH IN MESSIER 87

There are only a handful of BHs for which the Bondi accretion rate $\dot{M}_B$ or the mass inflow rate at the Bondi radius has been determined using observed temperature and density profiles. One of them is the AGN at the center of the galaxy Messier 87. It is one of the most closely studied nearby quasars. Di Matteo et al. (2003) determined its $\dot{M}_B$ at ~0.1 M☉/yr for an $M_{BH}$ =3x$10^9$ M☉. There are, however, several different estimates of $M_{BH}$ for M87; and hence Russell et al. (2015, and references therein) obtained a range of $\dot{M}_B$ from 0.1-0.2 M☉/yr for $M_{BH}$ =3.5-6.6x$10^9$ M☉. These measures of $\dot{M}_B$ provide a unique opportunity to further test the validity of the empirically derived Eq.3 and its applicability to the lowest-z BHs. To do so, however, we need an accurate estimate of radiative efficiency ε for the BH in M87 to convert the accretion rate $\dot{M}$ predicted by Eq.3 to the predicted mass-inflow rate $\dot{M}_T = \dot{M}/(1-ε)$. The foregoing results show that ε is a function of both z and $M_{BH}$. The M87 black hole has a redshift z=0.00428 and its mass is comparable to the masses of BHs in the green bin in Fig.3. The spline regression plot in Fig.3 shows that λ decreases with z, but reaches a near constant at z< 0.5 for BHs in the green bin. As discussed earlier, Kozlowski lists 35 quasars at z<0.5 in the green bin within a narrow z window of ~ 0.492-0.345 having a mean λ=0.0118. Substituting this value of λ in Eq.5, we get ε =0.84 for M87 at z <0.005 that is almost an order of magnitude higher than the commonly used canonical value of 0.1. Equation 3 gives $\dot{M}$ = 0.0176 - 0.033 M☉/yr for the range of $M_{BH}$ used by Russell et al. (2015) for the M87 black hole. Dividing this $\dot{M}$ by (1- ε) where ε=0.84, we get a predicted inflow rate $\dot{M}_T$= 0.11-0.21 M☉/yr that is essentially identical to the Bondi accretion rate of 0.1-0.2 $M_\odot$/yr (Russell et al. 2015) based on actual observational data and not on a scaling relation.

The agreement between predicted $\dot{M}_T$ and observed $\dot{M}_B$ for M87 is all the more remarkable in view of the fact that the empirically derived Eq.3 is based on the data of the highest-z SMBHs when the universe was < a billion years old, whereas M87 is almost as old as the universe. Furthermore, the empirical $\dot{M}_T$ is a function of a BH's $M_{BH}$ and ambient gas density, whereas the Bondi rate $\dot{M}_B$ is a function of $M_{BH}$, and gas density and temperature at the Bondi radius (see e.g. Russell et al., 2015). Why then the empirical relation accurately predict the $\dot{M}_B$ of the BH in M87 is an enigma. Also, there are several other BHs in nearby galaxies (NGC 3115, NGC 1600, and Cygnus A) and the Milky Way for which $\dot{M}_B$ have been derived using temperature and density profiles or other observational data. A paper in preparation compares the predicted $\dot{M}_T$ and reported $\dot{M}_B$ for all these BHs, finds a physical basis for the apparent enigma, and discusses the implications and/or constraints of the findings on accretion models.

## 7. CONCLUSIONS

We derived an empirical relation (Eq.3) defining a SMBH's spot/instantaneous accretion rate $\dot{M}$ as a function of its mass $M_{BH}$ and redshift z based on Eq.1, an empirical relation found by A22 using publicly available mass and age data for 59 high-z SMBHs and the commonly-used Salpeter relation. Equation 3 was applied to 20 quasars at z>5.3 and 11 PG quasars at z<0.5 to determine each BH's accretion rate $\dot{M}$ and the corresponding radiative efficiency ε. The empirical results were compared to their theoretical counterparts determined by TVN17 and DL11 using a scaling relation based on thin-disc accretion theory. Comparative tests showed that all empirical $\dot{M}$ and ε passed the tests, but the theoretical counterparts failed in most cases. The empirical mean value of ε for the 20 high-z quasars was determined to be ~0.26 or substantially higher than the commonly-used canonical value of 0.1

Based on Eq.3, a secondary empirical relation (Eq.5) was derived expressing the Eddington ratio λ as a function of z and ε. To further assess the validity of Eq.3 and the empirical mean value of ε ~0.26 for high-z quasars, we applied Eq.5 with ε=0.26 to 59 high-z BHs complied by A22 and compared the resulting distribution of λ to the observed distribution of λ for 50 high-z BHs determined by Shen et al. (2019). The two distributions of λ were found to be comparable and in good agreement.

The empirical relations imply that λ is an inverse function of $M_{BH}$ and decreases with z, and that ε is an inverse function of both $M_{BH}$ and z. We showed this to be the case by applying spline regression analysis to Kozlowski's (2017) catalog of λ as a function of z for 132,000 BHs at 0.1< z <2.4 separated into 3 BH mass bins (see Fig.3). In addition, Fig.4 illustrates the inverse dependence of ε on $M_{BH}$ for 34 high-z (>5.3) quasars with $M_{BH}$ >$10^8$ $M_\odot$. A least-squares fit to the data in Fig.4 is given by Eq.7 strictly applicable to BHs at z>5.7 and $M_{BH}$ >$10^8$ $M_\odot$.

The average λ for BHs >$10^9$ $M_\odot$ is determined to be ~0.25 at z >5.7 and ~0.012 at z <0.5. And the average ε for $M_{BH}$ >$10^9$ $M_\odot$ is found to be ~0.23 at z >5.7 and ~0.84 at z <0.005. The radiative efficiency ε of a BH at any redshift can be estimated using Eq.5 if its Eddington ratio λ is known. The uncertainty in thus determining a BH's ε depends upon the accuracy with which λ is determined.

We showed that the empirical mass-inflow rate (0.11-0.21 M☉/yr) onto the BH in M87 derived using Eq.3 and ε=0.84 matches its Bondi accretion rate (0.1-0.2 M☉/yr) determined by Russell et al. (2015) using observed density and temperature profiles.

In summary, the following are some of the insights into the growth and properties of SMBHs derived from the results of this study.

1. A BH's spot accretion rate $\dot{M}$ at any instant in its life span is a direct function of its mass and the ambient gas density. It increases as BH mass increases, but decreases with z in proportion to the decrease in the ambient gas density (Eq.3)
2. The spot accretion rate $\dot{M}$ initially increases exponentially from the inception of the seed at z~30, reaches a plateau from z~8.5 to 6, and steadily decreases thereafter (Eq.6 and Fig.2). Hence, the accretion time scale inferred from spot $\dot{M}$ is of little value and does not reflect the real accretion time of a BH
3. The previous findings 1 and 2 suggest that initially BH mass $M_{BH}$ or its gravitational reach increases faster than the decrease in gas density. The opposing effects of an increasing $M_{BH}$ and decreasing density reach a parity near z~7.5-7 and thereafter the effect of decreasing gas density supersedes the increase in $M_{BH}$.
4. At any redshift, larger SMBHs are apparently more efficient (lower radiative efficiency ε) in accreting gases than smaller BHs.
5. As z decreases towards zero, a SMBH becomes less efficient (higher ε) in accreting gases even though its mass keeps increasing, albeit at a decreasing rate; which suggests that ε may be an inverse function of gas density (see Eq.5).
6. At very low redshifts most of the available gas is radiated away by a BH and only a small fraction is accreted (e.g. M87).
7. The finding that ε varies from a low of ~013 at z~7.1 (BH#1, Table1) to a high of ~0.84 at z <0.005 applicable to the BH in M87 questions the usefulness of the wide spread use of a fixed value of ε=0.1.
8. The excellent agreement between the empirically determined mass inflow rate onto the BH in M87 and its "observed" Bondi accretion rate suggests that the mass inflow rate from the Bondi radius to the center of the BH apparently does not decrease significantly. More such examples are, however, needed to ascertain whether this is indeed the case in general.

Acknowledgements:

I thank Manuel Chirouze for his suggestion to apply spline analysis to the Kozlowski's data for black holes and for doing the plot shown in Fig.3.

**REFERENCES**

Aggarwal Y., 2022,  arXiv:2112.06338 **[astro-ph.GA]**

Bergström L., Goober I., 2006,  Cosmology and Particle Astrophysics, Springer 77

Bian W., Zhao Y.H., 2003, PASJ, 55, 59

Collin S., et al. 2002, A&A, 388, 771

Davis S. W., Laor A., 2011, ApJ, 728, 98

Di Matteo T., Allen S.W., Fabian A. C., Wilson A.S., Young A.J., 2003, ApJ, 582, 133–140

Kelly B.C., et al., 2010, ApJ, 719, 1315–1334

Kozłowski S., 2017, ApJ supplement, 228, 9, arXiv:1609.09489 (2017)

Netzer H., Trakhtenbrot B., 2014, MNRAS, 438, 672

Pacucci F., Loeb A., 2020, ApJ 895, 95

Planck group, 2020, A&A, 641, A6

Raimundo S. I., Fabian A. C., Vasudevan R. V., Gandhi P., Wu J., 2012, MNRAS, 419, 2529

Russell H.R., Fabian A.C., McNamara B.R., Broderick A.E., 2015, MNRAS, 451, 588-600

Salpeter, E. E., 1964, ApJ, 140, 796-800

Schulze A., Wisotzki L., 2010, arXiv: 1004.2671 [astro-ph.CO] (2010)

Shankar F., Crocce M., Miralda-Escudé J., Fosalba P., Weinberg D. H., 2010, ApJ, 718, 231

Shao Y., et al., 2017, ApJ, 845, 138

Shen Y., Greene J.E., Strauss M.A., Richards G.T., Schneider D.P., 2008, ApJ, 680, 169

Shen Y., et al., 2019, ApJ, 873, 35

Soltan A. 1982, MNRAS, 200, 115

Suh H., Hasinger G., Steinhardt C., Silverman J.D., Schramm M., 2015, ApJ, 815, 129

Trakhtenbrot B., Netzer H., 2012, MNRAS, 427, 3081

Trakhtenbrot B., Netzer H., Lira P., Shemmer O., 2011, ApJ, 730, 7

Trakhtenbrot B., Volonteri M., Natarajan P., 2017, ApJL, 836, L1

Wang F., et al., 2015, ApJL, 807, L9

Wickham H., 2016, ggplot2 Elegant Graphics for Data Analysis, Springer, Cham.

Willott C.J. et al., 2010, ApJ, 140, 546

Zubovas K., King A., 2021, MNRAS, 501, 4289